# Buoyancy waves in Pluto's high atmosphere: Implications for stellar occultations

W. B. Hubbard,[a,*] D. W. McCarthy,[b] C. A. Kulesa,[b] S. D. Benecchi,[c] M. J. Person,[d] J. L. Elliot,[d,e] and A. A. S. Gulbis[d,f]

[a] Lunar and Planetary Laboratory, The University of Arizona, 1629 E. University Blvd., Tucson AZ 85721, USA
*Corresponding Author E-mail address: hubbard@lpl.arizona.edu
[b] Steward Observatory, The University of Arizona, 933 N. Cherry Ave, Tucson, AZ 85721, USA
[c] Space Telescope Science Institute, 3700 San Martin Drive, Baltimore, MD 21218, USA
[d] Earth, Atmospheric and Planetary Sciences, MIT, 77 Massachusetts Ave., Cambridge, MA 02139, USA
[e] Physics, MIT, 77 Massachusetts Ave., Cambridge, MA 02139, USA
[f] Southern African Astronomical Observatory, P.O. Box 9, Observatory, 7935, Cape Town, South Africa

**Abstract**

We apply scintillation theory to stellar signal fluctuations in the high-resolution, high signal/noise, dual-wavelength data from the MMT observation of the 2007 March 18 occultation of P445.3 by Pluto. A well-defined high wavenumber cutoff in the fluctuations is consistent with viscous-thermal dissipation of buoyancy waves (internal gravity waves) in Pluto's high atmosphere, and provides strong evidence that the underlying density fluctuations are governed by the gravity-wave dispersion relation.

# 1. Introduction

The question of whether gravity waves cause the prominent scintillations ("spikes") frequently seen in stellar occultations by planetary atmospheres has been debated for over thirty years, dating from the availability of high-resolution observations of the 1971 occultation of β Scorpii by Jupiter (French and Gierasch 1974 [FG]; Veverka et al. 1974; Liller et al. 1974; Elliot et al. 1974). It is surprising that occultation observations of Pluto's tenuous and remote atmosphere can substantially clarify this matter.

The purpose of the present paper is to analyze the 2007 occultation of P445.3 (Person et al. 2008 [Paper I]; McCarthy et al. 2008 [Paper II]) from the point of view of gravity-wave and scintillation theory. Paper I discusses the multistation data set that furnishes the astrometric solution used in the present analysis, while Paper II provides details about the highest signal/noise data set, the dual-wavelength observations of a Pluto graze obtained at the southernmost station, the MMT Observatory, which forms the basis of the present analysis. See Papers I and II for observational and astrometric details of the P445.3 occultation, and Paper II for information about the best-fit isothermal-atmosphere model that we use to analyze the MMT data here.

This paper focuses on the prominent fluctuations in the observed stellar flux $\phi$ (where $\phi$ is normalized to unity outside of occultation). Using the astrometric solution presented in Paper I, we transform the observed $\phi(t)$, where $t$ = UTC of observation, to $\phi(r')$, where $r'$ is the distance of the observer from the star-Pluto vector in the shadow plane, a plane containing the observer and normal to the star-Pluto vector. Correspondingly, let $\phi_q(r')$ be the lightcurve for the (quiescent) background isothermal-atmosphere model, as described in Paper II and shown in Fig. 1 of that paper. The fluctuations which we analyze here are defined by $\Delta\phi = \phi - \phi_q$. We map each value of $r'$ onto a corresponding radius $r$ or $r_q$ via the relations

$$rdr/r'dr' = \phi \quad (1)$$

and

$$r_q dr_q/r'dr' = \phi_q \quad (2)$$

using the principle of conservation of flux and neglecting any contribution to the signal from the far-limb stellar image. Here $r$ is defined as the radius of closest approach to the center of Pluto for a refracted ray linking the star to the observer, and $r_q$ is the same quantity for refraction by the quiescent isothermal atmosphere. The closest-approach point is in the reference plane, a plane containing the center of Pluto and normal to the star-Pluto vector. Because the present event is always in the regime $0.7 \leq \phi_q \leq 1$, and because $|\Delta\phi| < 0.1$ in general (the maximum excursion is $\Delta\phi = -0.14$ in the visual; see below), differences between $r$ and $r_q$ are negligible for the purposes of this paper. Figure 1 shows the fluctuations in the two MMT channels, at full resolution. The fluctuations are well sampled and are observed with good signal/noise, a rarity for occultations observed in this range of $\phi_q$. The fluctuations are unusually broad from the point of view of scintillation theory, as they are much wider than a Fresnel scale (a circumstance discussed in detail in Section 3).

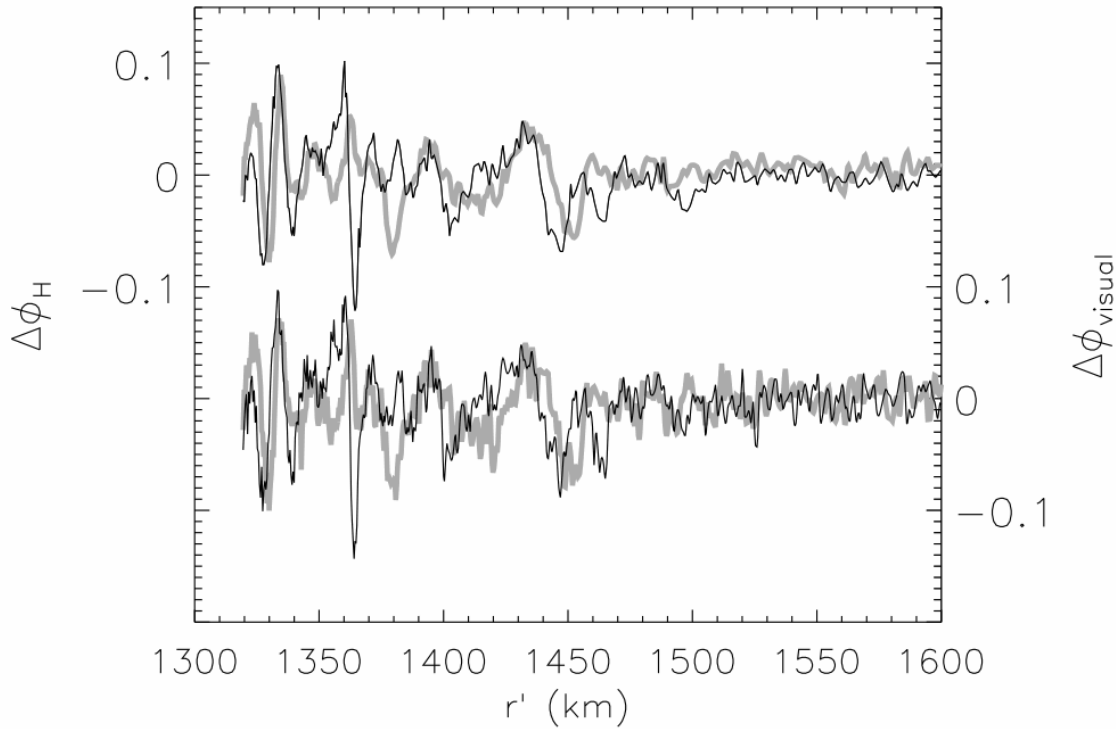

**Figure 1** – Lightcurve fluctuations at H-band (upper curves) and in the visual band (lower curves). Thin curves show immersion fluctuations (prior to midoccultation); thicker grey curves show emersion fluctuations (after midoccultation).

**2. Scintillation**

We use standard terminology of scintillation theory to discuss and analyze the statistical properties of the fluctuations shown in Fig. 1. But, although we use for convenience such terms as "scintillation" and "turbulence", we will present evidence that the broad fluctuations are produced by a physical mechanism that is very different from the scintillations that arise from light propagation through a scale-free (Kolmogorov) spectrum of density fluctuations in a turbulent medium. We make use of the analyses of Hubbard, Jokipii, and Wilking (1978; HJW) and French and Lovelace (1983; FL) and adhere as much as possible to their notation.

  

## 2.1. Parameters of the P445.3 occultation

Table 1 gives, for three representative values of $r_q$, chosen to span two equal intervals in radius, critical parameters for a scintillation analysis. The table also gives values for the Fresnel scale $\lambda_F = \sqrt{\bar{\lambda} D}$, where $\bar{\lambda}$ is the effective wavelength of the H-band and visual-band experiments respectively and $D$ is the distance to Pluto (see Table 1 and Papers I and II). Under the assumption that geometrical optics applies (i.e., there are no caustics or ray-crossing phenomena in the shadow) and that there is therefore a one-to-one mapping of $r'$ onto $r$ via Eq. (1), one may use an Abel integral transform pair (e.g., French et al., 1978) to invert $\phi(r')$ to obtain a profile of the atmospheric mass density $\rho(r)$. Let a horizontally-averaged density fluctuation be $\rho'(r) = \rho(r) - \bar{\rho}(r)$, where $\bar{\rho}(r)$ is the density profile for the average isothermal atmosphere. To $\bar{\rho}(r)$ there corresponds an average refractivity profile $\nu_A(r)$ and to $\rho'(r)$ there corresponds a turbulent refractivity fluctuation $\nu_t(r)$, in the notation of FL, and we have $\rho'(r)/\bar{\rho}(r) = \nu_t(r)/\nu_A(r)$. Upon integrating $\nu_A(r)$ through Pluto's atmosphere along a straight-line path with impact parameter $r$ and multiplying by $2\pi/\bar{\lambda}$, we obtain the total average phase shift $\Phi_A$. We obtain the random turbulent phase shift $\Phi_t$ by performing a similar operation on $\nu_t(r)$. Figure 2 shows results for the relative density fluctuations obtained from Abel inversions of the immersion and emersion H-band data. The overall correspondence of the immersion and emersion data indicates a high degree of layering of the density fluctuations.

**Table 1** – Scintillation parameters*

| $r_q$ (km) | $r'$ (km) | $\phi_q$ | $\Phi_A$ (H) | $\Phi_A$ (vis.) | $\Phi_t$ (H) | $\Phi_t$ (vis.) | $m_{scint}$ | $\lambda_{F, H}$ (km) | $\lambda_{F, vis.}$ (km) |
|---|---|---|---|---|---|---|---|---|---|
| 1448 | 1442 | 0.92 | 262 | 828 | 3.9 | 12.4 | 0.05 | 3.2 | 1.8 |
| 1398 | 1385 | 0.84 | 513 | 1620 | 2.6 | 8.1 | 0.06 | | |
| 1348 | 1319 | 0.70 | 1052 | 3326 | 5.3 | 16.7 | 0.07 | | |

*Phases are in radians.

Table 1 gives typical values of $\Phi_t$ obtained by using the inversion results shown in Fig. 2. We see that values of $\Phi_t$ are of order $2\pi$, meaning that the parameter $A$ of FL is large. However, the modulation index $m_{scint} = \sqrt{\langle (\Delta\phi/\phi_q)^2 \rangle} \ll 1$, so we are in the domain of weak scintillation, where atmospheric vertical scales and scintillation (shadow) vertical scales are comparable. As demonstrated by FL, under these circumstances the analytic weak-scintillation approximation made by HJW is valid, furnishing a powerful tool for analysis of the present data set.

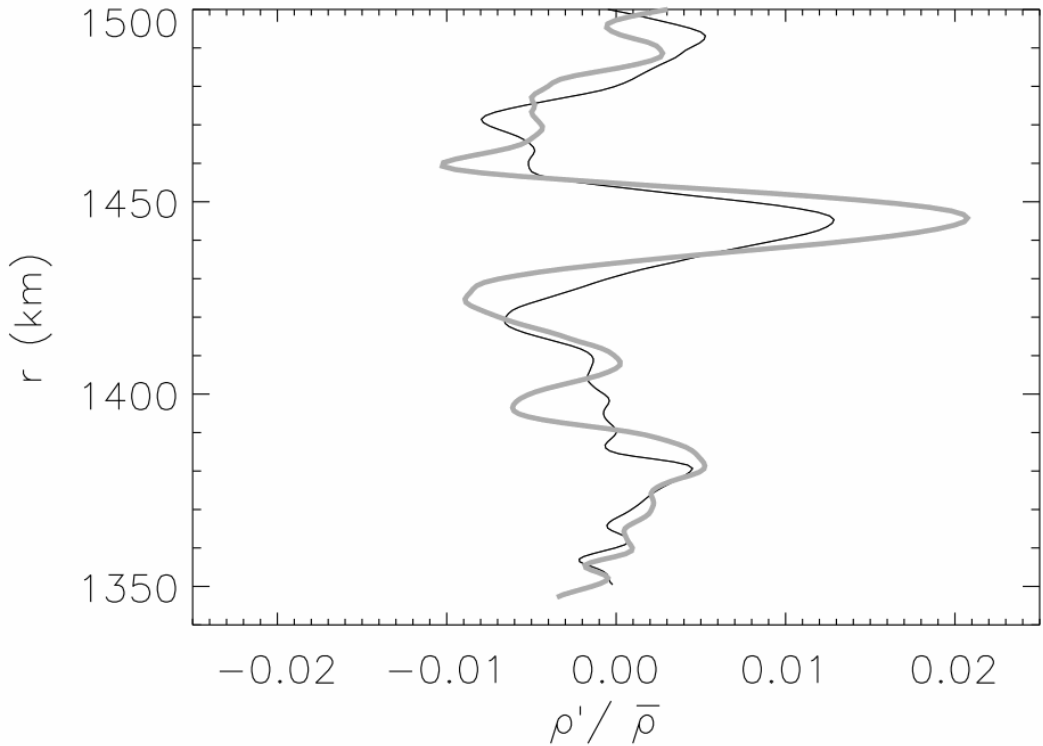

**Figure 2** – Inversions of the H-band lightcurve $\phi(r')$ showing the fluctuation in mass density $\rho'$ relative to the smooth (isothermal) mass density $\bar{\rho}$. The abscissa is also the ratio $v_t/v_A$ (see Section 2.1). Thin curve shows inversion of data prior to midoccultation; thicker grey curve shows inversion of data after midoccultation.

Figure 3 shows the logarithmic derivative $\nabla \equiv d\ln T/d\ln P$ from the same inversions, where $T(r)$ is the temperature and $P(r)$ is the pressure. The quantities plotted in Figs. 2 and 3 are insensitive to the large-scale offsets in the density that can arise from the choice of an arbitrary starting density for the inversion. The Schwarzschild stability criterion is

$$\nabla < (\gamma-1)/\gamma, \qquad (3)$$

where $\gamma = C_P/C_V = 1.4$ is the adiabatic index for $N_2$ gas, and $C_P$ and $C_V$ are respectively the heat capacities at constant pressure and constant volume. Figure 3 shows that the maximum values of $\nabla$ are about a factor of two smaller than the stability limit.

 

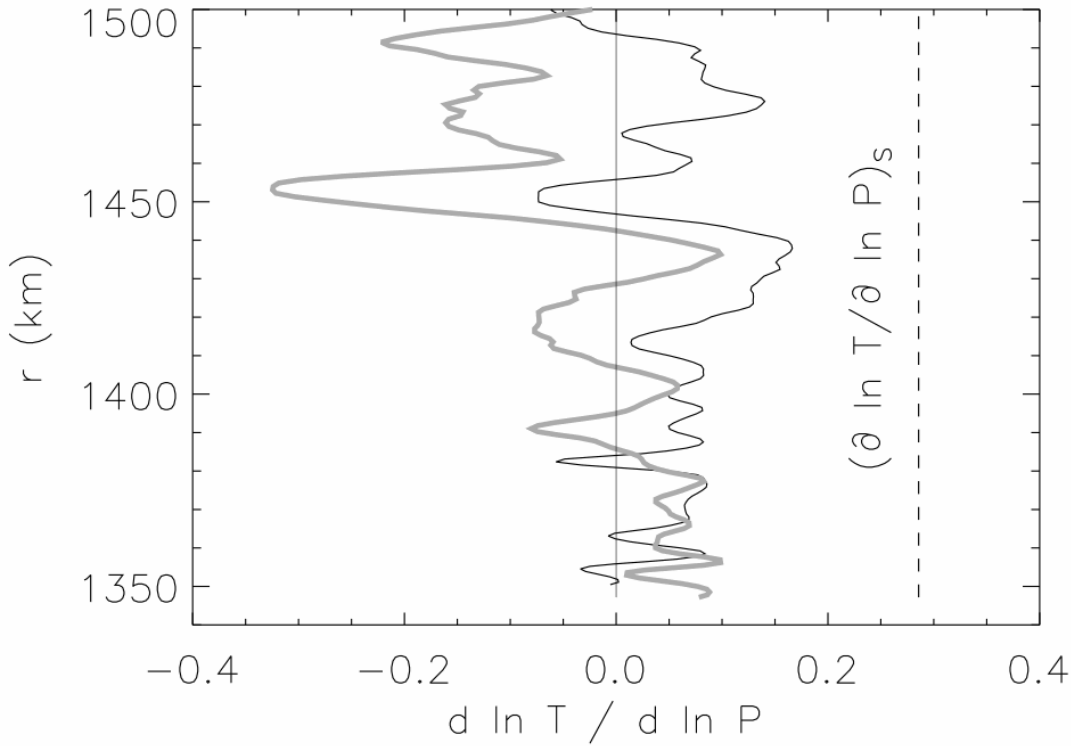

**Figure 3** – Logarithmic temperature-pressure gradient from lightcurve inversions. The thin curve shows inversion of immersion data and the thicker grey curve shows inversion of emersion data. Dashed line shows the adiabatic gradient.

2.2. Statistical models for scintillation

Following HJW, we start with $P_v(k, l, m)$, the three-dimensional power spectrum for refractivity fluctuations in Pluto's atmosphere. Here $m$ is the wavenumber of a fluctuation's variation perpendicular to Pluto's limb, $k$ is the wavenumber for the variation parallel to the limb, and $l$ is the wavenumber for the variation along the line of sight, all from the perspective of the observer. HJW considered an anisotropic spectrum,

$$P_v(k, l, m) \propto (\tilde{\rho}^2 k^2 + \tilde{\rho}^2 l^2 + m^2)^{-\alpha} \qquad (4)$$

where $\tilde{\rho}$ is an anisotropy parameter and $\alpha = 11/6$ for three-dimensional Kolmogorov turbulence (however, as we show in Section 3, the Kolmogorov index turns out to be inconsistent with this Pluto data set). As shown by HJW, the two-dimensional power spectrum $P_p(k, m)$ for $\Phi_t$, the phase fluctuations of the photon signal after propagation through Pluto's atmosphere, has the functional form of Eq. (4) with $l = 0$. The value of $\tilde{\rho}$ cannot be precisely determined from the present data set, but we know that $\tilde{\rho} \gg 1$ because of the high degree of symmetry of immersion vs. emersion fluctuations (Fig. 1).

For an alternative model of the dynamics of Pluto's upper atmosphere, we turn for guidance to the Earth's atmosphere at comparable pressure levels. According to Fritts (1984) and Smith et al. (1987), density fluctuations at these levels are largely the manifestation of a saturated spectrum of gravity (buoyancy) waves. The amplitude of the saturated power spectrum is limited by condition (3), and as discussed by Smith et al. (1987), is typically a factor two or three below this limit because of wave superposition. The one-dimensional (vertical) power spectrum is obtained by integrating (4) over all values of $k$ and $l$. The result of this integration varies as $m^{-3}$ for the Smith et al. spectrum, while the corresponding Kolmogorov vertical power spectrum is much less steep, varying as $m^{-5/3}$. The Smith et al. model includes a low-$m$ cutoff $m_*$ which slowly decreases with altitude. To test the Smith et al. model, we adopt

$$P_\text{p}(k,m) \propto (m_*^2 + \tilde{\rho}^2 k^2 + m^2)^{-\alpha} \tag{5}$$

with $\alpha = 5/2$.

Following HJW and FL, we compute the autocorrelation function (ACF) of the fluctuations in the stellar flux:

$$C(\Delta x, \Delta y) = K \int_{-\infty}^{\infty} dm \int_{-\infty}^{\infty} dk\, e^{ik\Delta x + im\Delta y} \sin^2[(k^2 + m^2/\phi_q)(\lambda_F^2/4\pi)] \left(m_*^2 + \tilde{\rho}^2 k^2 + \frac{m^2}{\phi_q^2}\right)^{-\alpha}, \tag{6}$$

where $\Delta x, \Delta y$ is the spatial offset (parallel, perpendicular) to the limb in the shadow plane and $K$ is a normalizing constant such that $C(0,0) = 1$. To find $C$ in the limit $\tilde{\rho} \to \infty$, rewrite (6) with the substitution $\kappa = \tilde{\rho} k$, leading to

$$\begin{aligned} C(\Delta x, \Delta y) &= K \int_{-\infty}^{\infty} dm \int_{-\infty}^{\infty} d\kappa\, e^{\frac{i\kappa\Delta x}{\tilde{\rho}} + im\Delta y} \sin^2[(\kappa^2/\tilde{\rho}^2 + m^2/\phi_q)(\lambda_F^2/4\pi)] \left(m_*^2 + \kappa^2 + \frac{m^2}{\phi_q^2}\right)^{-\alpha} \\ &\approx K \int_{-\infty}^{\infty} dm\, e^{im\Delta y} \sin^2[(m^2/\phi_q)(\lambda_F^2/4\pi)] \int_{-\infty}^{\infty} d\kappa \left(m_*^2 + \kappa^2 + \frac{m^2}{\phi_q^2}\right)^{-\alpha} \\ &= K' \int_{-\infty}^{\infty} dm\, e^{im\Delta y} \sin^2[(m^2/\phi_q)(\lambda_F^2/4\pi)] \left(m_*^2 + \frac{m^2}{\phi_q^2}\right)^{-\beta}, \end{aligned} \tag{7}$$

where $K'$ is another normalizing constant and $\beta = 2$ for a spectrum of saturated gravity waves, while $m_* = 0$ and $\beta = 4/3$ for highly anisotropic Kolmogorov turbulence. For comparison with our Pluto data, we use the predicted one-dimensional power spectrum of the fluctuations of the stellar flux in the shadow plane, normal to the projected limb, which is just the Fourier transform of the one-dimensional ACF, or

$$P_\phi(m) \propto \sin^2[(m^2/\phi_q)(\lambda_F^2/4\pi)]\left(m_*^2 + \frac{m^2}{\phi_q^2}\right)^{-\beta}. \tag{8}$$

## 3. Waves in Pluto's high atmosphere

3.1. Theory

FG carried out the first systematic analysis of possible manifestations of atmospheric waves in planetary occultations. They considered the dispersion relations for internal gravity (buoyancy) waves of angular frequency $\omega$ and horizontal and vertical wavenumbers $k$ and $m$,

$$\omega^2 = \frac{\omega_B^2 k^2 + f^2 m^2}{k^2 + m^2}, \tag{9}$$

where $\omega_B$ is the buoyancy frequency (the value of $\tau_B = 2\pi/\omega_B = 1.2$ hours is given in Table 2) and $f$ is the coriolis frequency ($2\pi/f = 153$ hours at 30 degrees latitude), and for Rossby waves,

$$\omega = \frac{\beta_c k}{k^2 + f^2 m^2 \omega_B^{-2}}, \tag{10}$$

where $\beta_c \sim 0.05$ km$^{-1}$hr$^{-1}$ is the first derivative of $f$ with respect to northward displacement. We calculate $f$ for a representative latitude of 30 degrees (the actual occultation covered a broad range of Pluto latitudes, see Paper I; values of $\tau_B = 2\pi/\omega_B$ and $\tau_C = 2\pi/f$ are given in Table 2). In the data analysis presented below, we do include the coriolis term in the dispersion relation (9). However, for the values of $m$ of interest, the term in $f$ has only a small effect due to the slowness of Pluto's rotation and the low latitudes of the occultation points on Pluto, so the limiting form of Eq. (9), $\omega = k\omega_B/m$ is a reasonable approximation.

For either of the two putative dispersion relations, the wavenumber $k \approx 2\pi/1000$ km$^{-1}$ (see Paper I) is fixed by the geometry, while $m$ is measured. Therefore, if one could determine the third variable $\omega$, one (or neither) of relations (9), (10) could be selected. Meaningful limits can be placed on $\omega$ by considering dissipation of the waves through diffusion of momentum and heat, as was originally done by FG. The relevant molecular transport coefficients are the kinematic viscosity $\nu_V = \eta/\rho$, where $\eta$ is the molecular viscosity of dilute N$_2$ gas at $T = 100$ K and $\rho$ is the mass density, and the thermal diffusivity $\chi = K_{\text{cond}}/\rho C_P$, where $K_{\text{cond}}$ is the thermal conductivity. Table 2 gives values of $\nu_V$ and $\chi$ (Lemmon and Jacobsen 2004) at three atmospheric levels relevant to this occultation.

**Table 2** – Mean-atmosphere parameters

| $r_q$ (km) | $P$ (Pa) | $\rho$ (kg/m³) | $H$ (km) | $\nu_V$ (m²/s) | $\chi$ (m²/s) | $\tau_B$ (h) | $\tau_C$ (h) |
|---|---|---|---|---|---|---|---|
| 1448 | 0.015 | $5.2\times10^{-7}$ | 71 | 13 | 17 | 1.2 | 153 |
| 1398 | 0.032 | $1.1\times10^{-6}$ | 66 | 6 | 8 | | |
| 1348 | 0.070 | $2.3\times10^{-6}$ | 62 | 3 | 4 | | |

Landau and Lifshitz (1959) derive for the damping of a sound wave in a gas with wavenumber $m$

$$(\dot{E}/\bar{E})_{sound} = -m^2\left[\frac{4}{3}\nu_V + \frac{\zeta}{\rho} + \frac{1}{2}(\gamma-1)\chi\right], \qquad (11)$$

where $\dot{E}$ is the rate of dissipation of wave energy per unit volume and $\bar{E}$ is the average wave energy per unit volume. In Eq. (11), $\zeta$ is the bulk viscosity. As might be expected, the expression for damping of a highly anisotropic gravity wave with $m \gg k$ resembles Eq. (11), but with different coefficients multiplying the transport coefficients. Following the analysis of Landau and Lifshitz, we obtain

$$(\dot{E}/\bar{E})_{gravity\ wave} \equiv -\tau_m^{-1} = -m^2\left[2\nu_V + \frac{5}{2}(\gamma-1)\chi\right], \qquad (12)$$

where we neglect the contribution from bulk viscosity as gravity waves are approximately incompressible.

Define an amplitude factor giving the decrease in wave energy after one oscillation,

$$A_m = \exp(-\tau/\tau_m), \qquad (13)$$

where $\tau = 2\pi/\omega$ is the wave period. If $A_m \approx 1$, a wave can persist for more than one cycle. Waves with $A_m \ll 1$ will not be seen.

3.2. Comparison with data

For an initial comparison of theoretical with observed $P_\phi(m)$, we obtain the Fourier transform and thus the power spectrum of $\Delta\phi(r')$ on the shadow plane over the interval $1319 \le r' \le 1442$ km, corresponding to a 100-km interval in $r_q$ (Table 1; note that the $\Delta\phi$ power spectra of Paper II are computed on the reference plane using the mapping $r' \to r_q$ and therefore differ slightly from those presented here). Data points shown in Fig. 4 are obtained by averaging four individual power spectra (immersion and emersion in H and visual), and the error bar on each data point is derived from the dispersion of the four individual power spectra about the mean. In Figs. (4-6), open data

                                                                                                p. 9

points at $m = 0.1$ km$^{-1}$ are not meaningful because the corresponding wavelength is approximately a scale height.

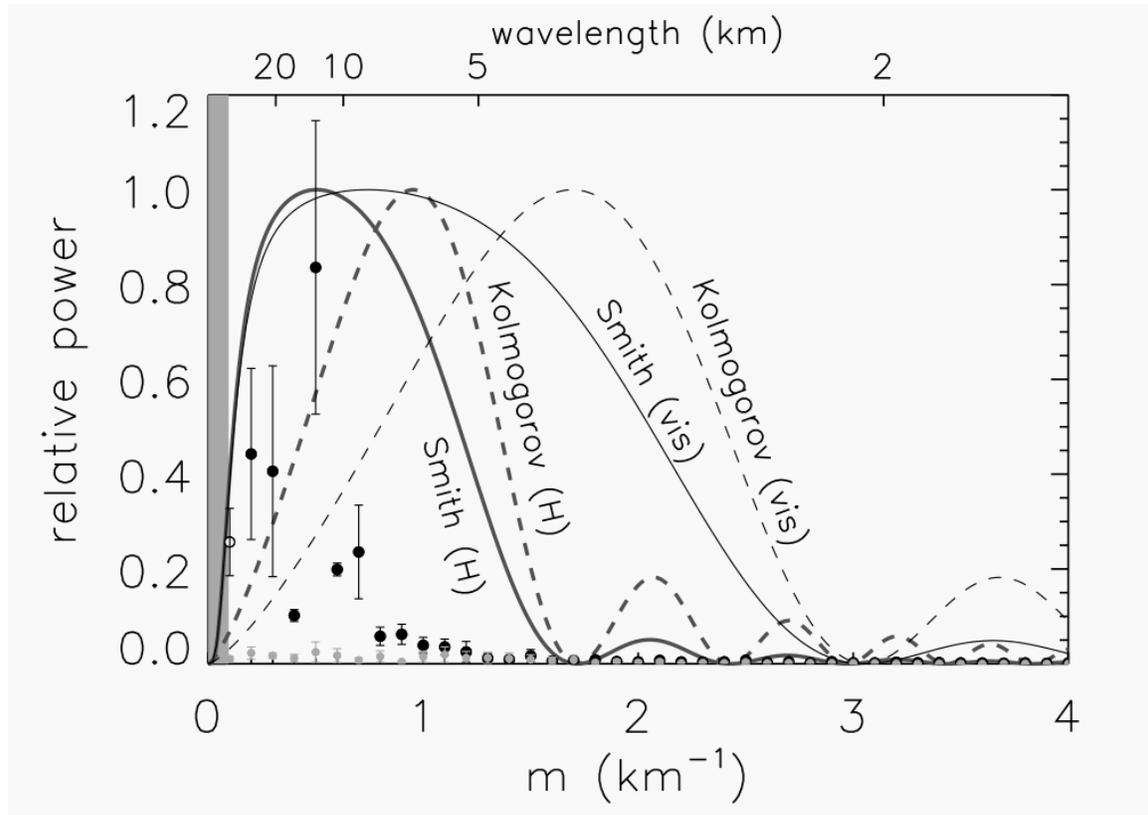

**Figure 4** – Theoretical (curves) and observed (solid data points) power spectra as a function of vertical wavenumber $m$; top scale shows the corresponding vertical wavelength $2\pi/m$. Grey data points indicate observed pre-immersion and post-emersion power spectrum and thus show the contribution of terrestrial scintillation, which dominates any Pluto contribution for $m > 1$ km$^{-1}$. Grey shaded vertical band denotes vertical wavelengths greater than the scale height $H$. Solid curves are calculated spectra (H-band and visual-band) for the saturated gravity-wave model of Smith *et al.* (1987). Dashed curves assume a Kolmogorov spectrum with no outer scale. In this figure, the observed power spectrum is computed over the shadow-plane radius interval $1319 \leq r' \leq 1442$ km, corresponding to a reference-plane radius range of 100 km, or $1348 \leq r_q \leq 1448$ km, essentially spanning all observed fluctuations.

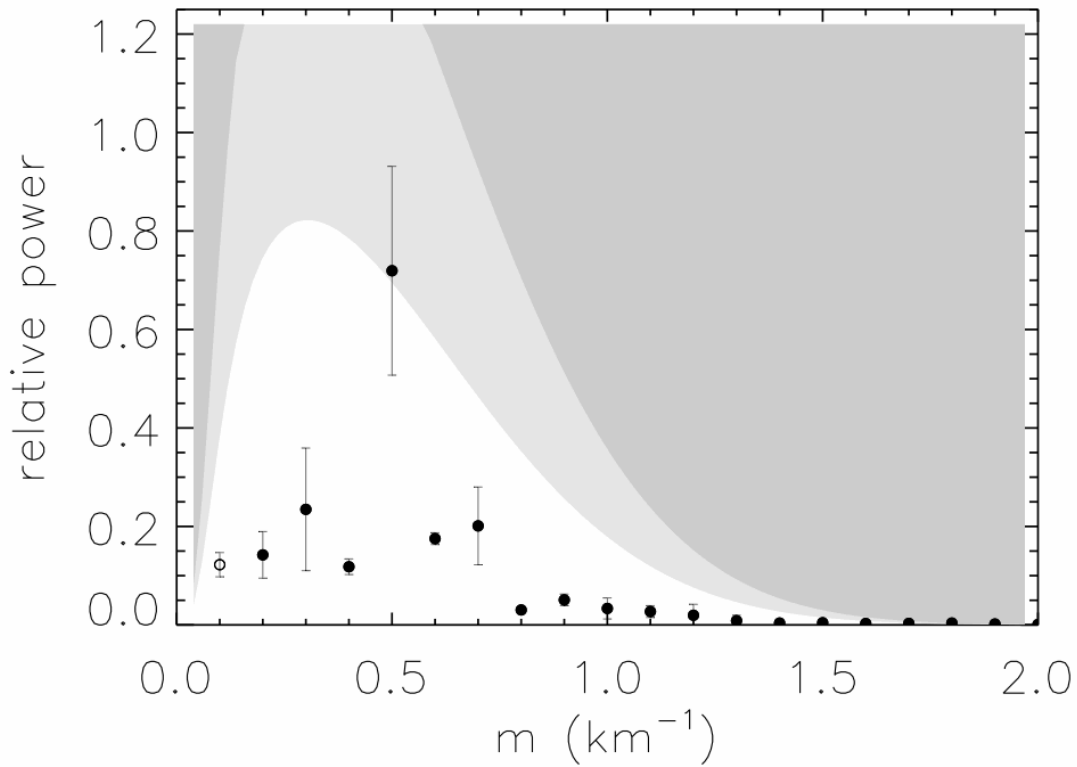

**Figure 5** – Data points showing the observed power spectrum for the lowest 50-km interval in $r$ (1348 to 1398 km in $r$ or 1319 to 1385 km in $r'$). The grey forbidden area is delimited below by the theoretical gravity-wave spectrum of Fig. 4 multiplied by $A_m$ as predicted by dispersion relation (9) – see text for a discussion of the two grey areas. At wavenumbers greater than ~ 0.7 km$^{-1}$, the factor $A_m$ cuts off oscillations within less than one cycle. Because of the high-$m$ cutoff, there is no significant difference between the theoretical H-band and visual-band grey areas.

  

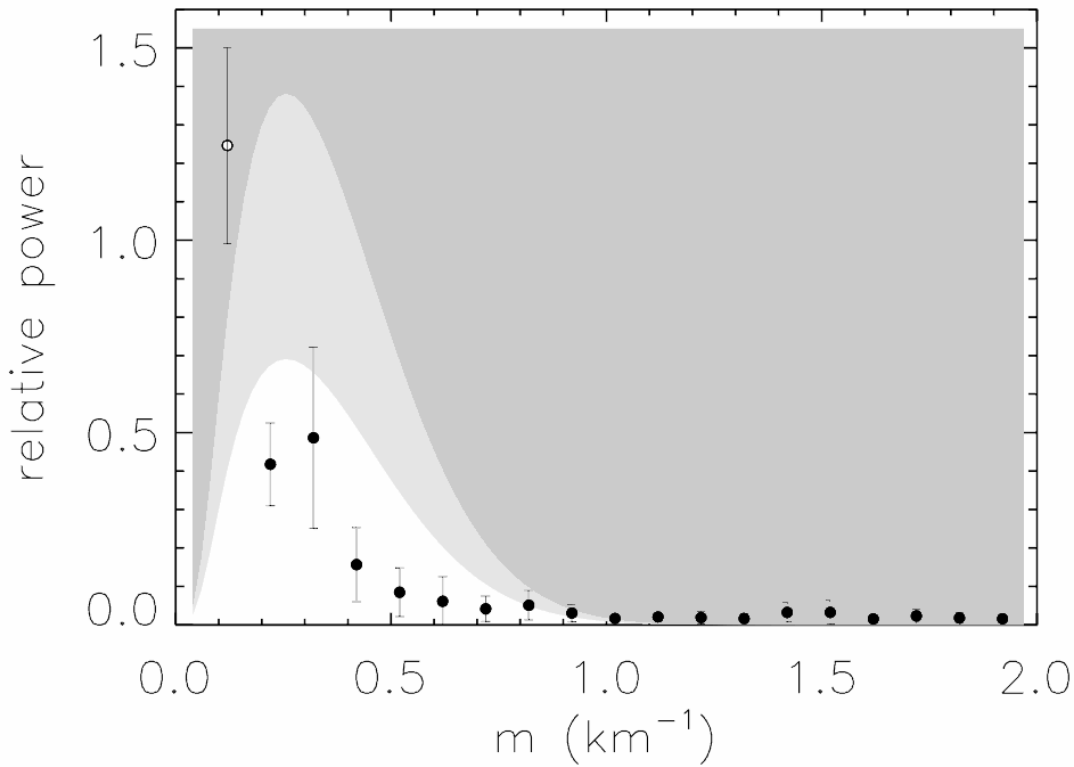

**Figure 6** – Same as Fig. 5, but for the uppermost 50 km in $r$. Again, the open data point at $m = 0.1$ km$^{-1}$ is not meaningful; any power that is actually at this wavelength could perhaps represent Rossby waves rather than buoyancy waves.

The relative vertical scales in Fig. 4 cannot be precisely determined; for convenience we normalize the theoretical and observational spectra to the same peak power (allowing for the error bar of the peak data point). We have seen that the peak observed density fluctuations are roughly a factor 2 below the maximum amplitude for monochromatic wavebreaking, or roughly a factor 3 to 4 below the corresponding power-spectrum point. According to Smith et al. (1987), the theoretical power spectrum should also have an amplitude about a factor 3 below the wavebreaking limit. Our observed power spectrum has a few isolated peaks at discrete $m$ rather than a continuous spectrum, and even at low $m$ it does not fill the envelope predicted by the Smith et al. spectrum. Although we can rule out a small spectral index (such as Kolmogorov), we cannot confirm a specific spectral index such as Smith. We believe that this uncertainty is due to inadequate statistics. Our single pass through Pluto's atmosphere could detect only a finite number of waves from the ensemble of possibilities.

According to the theoretical predictions shown in Fig. 4, for either a saturated gravity-wave spectrum ($\beta = 2$) or for a Kolmogorov spectrum ($\beta = 4/3$), we should have seen significant power at $m > 0.8$ km$^{-1}$. The combination of our instrumental cadence and the grazing geometry would have allowed detection of power up to at least $m \sim 4$ km$^{-1}$. The high-$m$ cutoff imposed by wave optics (the so-called Fresnel filter function $\sin^2[(m^2/\phi_q)(\lambda_F^2/4\pi)])$ lies well above the observed $m$ cutoff in the data. Thus, the high-$m$ cutoff is not a wave-optical phenomenon. At the same time, the Kolmogorov spectrum is also cut off by the Fresnel filter function at low $m$, while the steeper gravity-wave spectrum is only cut off at low $m$ by the cutoff wavenumber $m_*$. Because of the limited radius range spanned by our $\Delta\phi(r')$ data, we cannot determine $m_*$ or even if there is any low-$m$ cutoff; the solid theoretical curves in Fig. 4 assume $m_* = 2\pi/60$ km$^{-1}$. However, at low $m$ the data in Fig. 4 are clearly more consistent with the saturated gravity-wave spectrum than they are with a Kolmogorov spectrum.

Data shown in Fig. 5 plot the fluctuation power over the range $1319 \leq r' \leq 1385$ km, corresponding to a reference-plane radius $1348 \leq r_q \leq 1398$ km, the lowest 50 km of the data set. Grey areas show ranges where, according to theory, no power should be seen. The boundary of the lowest grey area is defined by the product of the Smith et al. spectrum and the suppression factor $A_m$ defined by Eq. (13). This boundary is the same for the H band and the visual band because $A_m \ll 1$ where the H-band and visual-band Smith *et al.* spectra are different. The vertical scale of the ordinate is determined by requiring the lowest grey boundary to pass near the highest data point, and is in fact consistent with the normalizations of Fig. 4. The boundary between the dark-grey and light-grey regions is a factor of two higher in power, and depicts a reasonable uncertainty in determining the upper boundary. We stress that the high-$m$ cutoff due to $A_m \ll 1$ comes into play well before the Fresnel filter function cutoff, which is why we have the robust result that there is no wavelength dependence in the observed power.

Figure 6 shows fluctuation power for the range $1385 \leq r' \leq 1442$ km, corresponding to a reference-plane radius range spanning $1398 \leq r_q \leq 1448$ km, the highest range for which we can measure scintillation power. The ordinate scales (theory and data) are unchanged from Fig. 5, showing that the observed scintillation power is lower. Transport coefficients for this height interval are larger, causing the theoretical cutoff due to $A_m$ to move to lower $m$, and the position of the theoretical maximum power to move lower as well, consistent with the data.

To test whether results presented in Figs. 5 and 6 are sensitive to the chosen altitude bins, we present the scintillation-power analysis in a different way in Fig. 7. To obtain adequate statistics, we use an altitude bin of 60 km width (about a scale height), and we slide this bin from the deepest penetration point in the shadow (corresponding to a central bin position of $r' = 1349$ km) to the largest radius where scintillations can still be measured (corresponding to a central bin position of $r' = 1418$ km). Figure 7 is a contour plot showing how the power spectrum changes with altitude, and is consistent with our interpretation of Figs. 5 and 6.

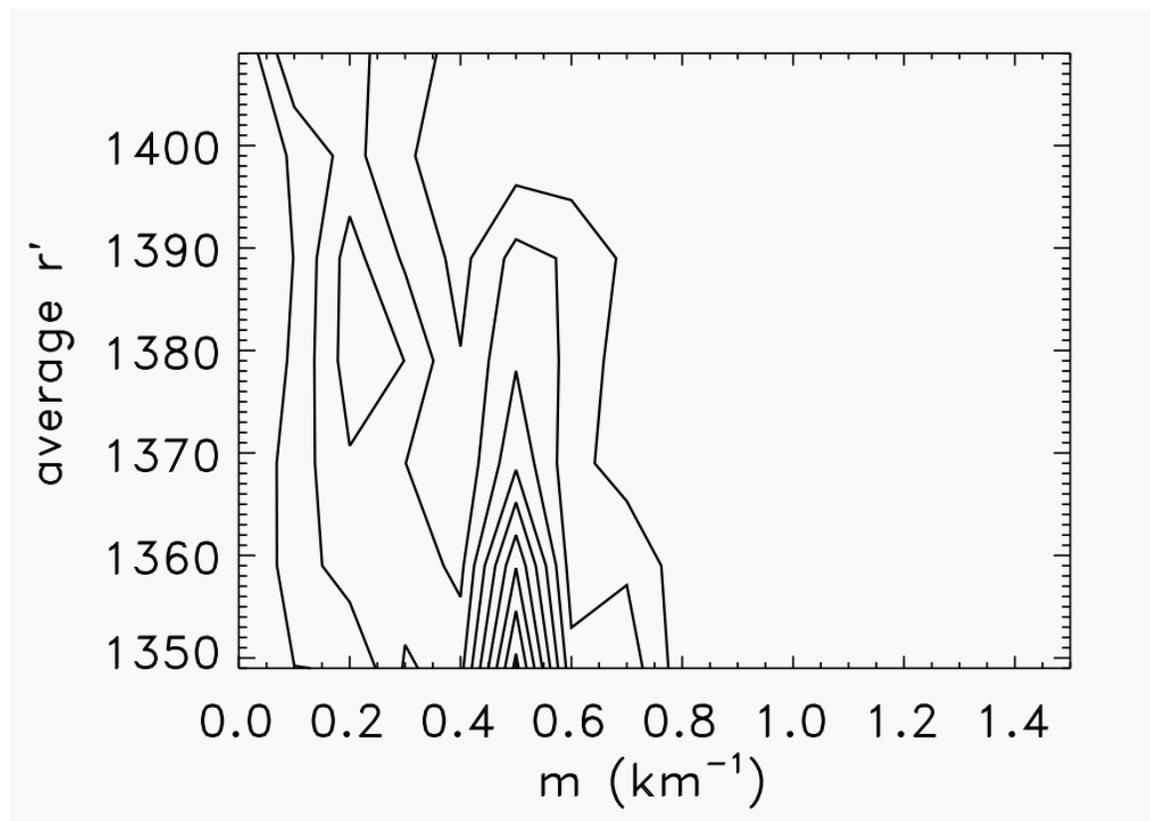

**Figure 7** – An alternative representation of results shown in Figs. 5 and 6, this time as a contour plot of scintillation power versus average distance from the center of Pluto's shadow. About ten equally-spaced contours of constant power are shown.

**4. Conclusions**

4.1 Implications and predictions

The saturated gravity-wave spectrum proposed for the Earth is consistent with our Pluto data, when suitably scaled. This circumstance suggests the possibility of a general model for scintillations in occultations. The model presented here could be used to predict the properties of the scintillations in other occultation data sets, and a number of published scintillation studies could be revisited. As a modest first step, we show in Fig. 8 a weak-scintillation prediction of the expected fluctuation spectrum for Pluto for an occultation that probes 50 km deeper in $r$ than the present one, corresponding roughly to $\phi_q \approx 1/2$ at deepest penetration. This height reduction amounts to nearly one density scale height, causing an e-fold decrease in transport coefficients (see Table 2), allowing buoyancy waves of larger $m$ values to persist. We thus predict higher spatial frequencies based on the HJW theory, and if the strong-scintillation predictions of FL are applicable to this lower value of $\phi_q$, even higher frequencies might be detected.

           

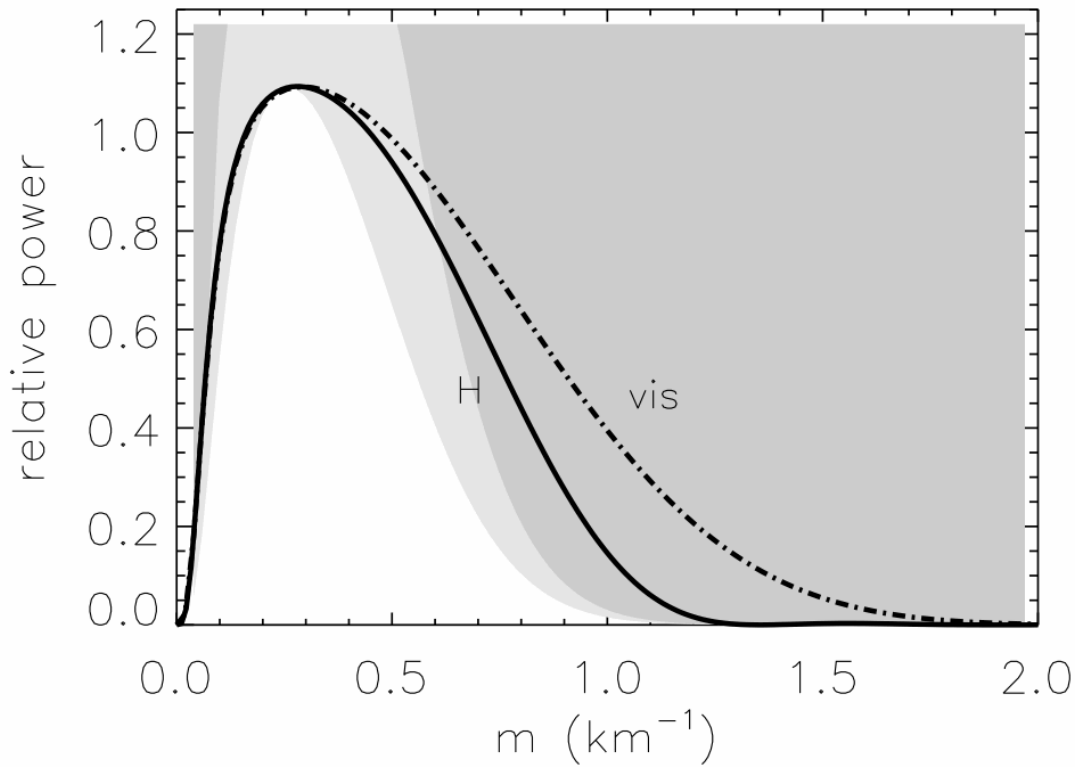

**Figure 8** – Theoretical prediction of scintillation power spectra (heavy solid line: H-band; dot-dashed line: visual-band) at a deeper level in Pluto's atmosphere, $r = 1298$ km, corresponding to $\phi_q \approx 1/2$. For H-band the Fresnel cutoff at high $m$ would begin to be as important as the diffusion cutoff. A visual-band occultation should therefore observe more scintillation than an H-band occultation, at high spatial wavenumbers (corresponding to spatial wavelengths < 5 km). For comparison, the grey areas repeat the predicted power spectrum (with uncertainty in vertical scale) shown in Fig. 5 for a level ~50 km higher in Pluto's atmosphere.

A data set from a 1999 Jupiter occultation, somewhat analogous to the present one because it pertains to grazing incidence through high atmospheric layers, is analyzed by Raynaud *et al.* (2003). Importantly, the power spectrum of density fluctuations inferred in that paper corresponds to the Smith *et al.* law. But despite multistation data, the profiles of Jupiter's sub-microbar atmosphere are not of as high spatial resolution or signal/noise ratio as the present Pluto data set. Raynaud *et al.* found evidence in their data set for an individual Jovian density-fluctuation wave with $2\pi/m \approx 3$ km and $2\pi/k \approx 70$ km, but they concluded that it could not be a buoyancy wave because of dissipation considerations. Perhaps a reanalysis of Jupiter data from a broader statistical point of view would lead to different conclusions.

4.2 Why this Pluto occultation data set is exceptional

We claim to have demonstrated the presence of buoyancy (internal-gravity) waves in Pluto's upper atmosphere, by means of an unambiguous determination of $m, k, \omega$ points on the gravity-wave dispersion relation using the mapping of an $m$–cutoff onto a timescale ($\omega$) axis. For the sake of argument, let us hypothesize that the Pluto scintillations are caused by static layered structures having nothing to do with gravity waves. This assumption would leave the observed $m$–cutoff unexplained. Large-scale static structures would take longer to dissipate than small-scale structures, but apart from that we would have no physics to connect the $m$–cutoff to dissipation timescales.

Assuming that we are seeing buoyancy waves, one can then place an upper limit on the energy flux carried by the waves, as follows. According to Fritts (1984), a gravity wave "breaks" (goes unstable in the sense of violating inequality [3]) when the wave's horizontal velocity amplitude $u'$ exceeds the wave's horizontal phase speed $c = \omega/k$. The corresponding maximum energy flux is then given by $J_{\max} \approx c \frac{1}{2}\rho c^2$, and, using the limiting form of Eq. (9), $\omega = k\omega_B/m$, we find that $J_{\max} \propto m^{-3}$. Considering a monochromatic gravity wave with $m \approx 0.5$ km$^{-1}$ (vertical wavelength about 10 km) and $k \ll m$, we find $J_{\max} \approx 4 \times 10^{-8}$ watt/m$^2$, many orders of magnitude smaller than the solar constant at Pluto (about 1 watt/m$^2$). We consider this to be a reasonable value.

For this event, the relevant $m, k, \omega$ points are within a detectable range because of a convergence of circumstances: atmospheric frequencies are very low for Pluto because of its low gravity and slow rotation, while at the same time, the grazing occultation by the highest levels of Pluto's atmosphere makes it possible to probe, with high spatial resolution and signal/noise, layers in its atmosphere where pressures are well below a microbar and transport coefficients are therefore very large. A further advantage provided by Pluto is that despite the brightness of P445.3 relative to Pluto, in absolute terms it is a faint star with small angular diameter, much smaller than a Fresnel scale when projected over Pluto. Our experiment is thus sensitive to small spatial scales < 3 km, whose absence is a critical piece of evidence. Finally, the grazing nature of the occultation limits us to a range of $\phi_q$ where a straightforward weak-scintillation theory is applicable.

Are some of the waves in the Pluto data actually Rossby waves obeying dispersion relation (10), as suggested in Paper I? For the observed range of $m$, the corresponding $\omega$ values would be low compared with those of gravity waves, and FG have argued that Rossby waves might dissipate too rapidly in general. Nevertheless, dissipation timescales for very low $m$ Rossby waves, those with wavelengths on the order of a Pluto scale height, can exceed their oscillation periods. We cannot rule out a contribution by such Rossby waves to the power spectrum, although they are unlikely to be a major component.

**Acknowledgments**

Observations reported here were obtained at the MMT Observatory, a joint facility of the University of Arizona and the Smithsonian Institution. We acknowledge the extensive engineering support of the entire MMTO staff, especially Mr. Shawn Callahan. This publication makes use of data products from the Two Micron All Sky Survey, which is a joint project of the University of Massachusetts and the Infrared Processing and Analysis Center/California Institute of Technology, funded by the National Aeronautics and Space Administration and the National Science Foundation. This work was partially funded by NASA Planetary Astronomy grants NNG04GE48G and NNG04GF25G. Partial funding for MMTO observations was also provided by Astronomy Camp. We thank two referees for helpful comments.

**References**

Elliot, J. L., Wasserman, L. H., Veverka, J., Sagan, C. Liller, W. 1974. The occultation of beta Scorpii by Jupiter. II. The hydrogen-helium abundance in the jovian atmosphere. Astrophys. J. 190, 719-730.

French, R. G., Elliot, J. L., Gierasch, P. J. 1978. Analysis of stellar occultation data. Icarus 33, 186-202.

French, R. G., Gierasch, P. J. 1974. Waves in the jovian upper atmosphere. J. Atm. Sci. 31, 1707-1712 (FG).

French, R. G., Lovelace, R. V. E. 1983. Strong turbulence and atmospheric waves in stellar occultations. Icarus 56, 122-146 (FL).

Fritts, D. C. 1984. Gravity wave saturation in the middle atmosphere: A review of theory and observations. Rev. Geophys. and Space Phys. 22, 275-308.

Hubbard, W.B., Jokipii, J.R., Wilking, B.A. 1978. Stellar occultations by turbulent planetary atmospheres: A wave-optical theory including a finite scale height. Icarus 34, 374-395 (HJW).

Landau, L.D., Lifshitz, E.M. 1959. Fluid Mechanics. Addison-Wesley: Reading, MA, Sections 49 and 77.

Lemmon, E. W., Jacobsen, R. T. 2004. Viscosity and thermal conductivity equations for nitrogen, oxygen, argon, and air. Intl. J. Thermophys. 25, 21-69.

Liller, W., Elliot, J. L., Veverka, J., Wasserman, L. H., Sagan, C. 1974. The occultation of beta Scorpii by Jupiter. III. Simultaneous high time-resolution records at three wavelengths. Icarus 22, 82-104.

McCarthy, D. W., Hubbard, W. B., Kulesa, C. A., Benecchi, S. D., Person, M. J., Elliot, J. L., Gulbis, A. A. S. 2008. Long-wavelength density fluctuations resolved in Pluto's high atmosphere. Astron. J. 136, 1519-1522 (Paper II).

Person, M. J., Elliot, J. L., Gulbis, A. A. S., Zuluaga, C. A., Babcock, B. A., McKay, A. J., Pasachoff, J. M., Souza, S. P., Hubbard, W. B., Kulesa, C. A., McCarthy, D. W., Benecchi, S. D., Levine, S. E., Bosh, A. S., Ryan, E. V., Ryan, W. H., Meyer, A., Wolf, J., Hill, J. 2008. Waves in Pluto's upper atmosphere. Astron. J. 136, 1510-1518 (Paper I).

Raynaud, E., Drossart, P., Matcheva, K., Sicardy, B., Hubbard, W. B., Roques, F., Widemann, T., Gladstone, G. R., Waite, J. H., Nadeau, D., Bastien, P., Doyon, R., Hill, R., Rieke, M. J., Marley, M. 2003. The 10 October 1999 HIP 9369 occultation by the northern polar region of Jupiter: ingress and egress lightcurves analysis. Icarus 162, 344-361.

Smith, S.A., Fritts, D.C., VanZandt, T.E. 1987. Evidence for a saturated spectrum of atmospheric gravity waves. J. Atmosph. Sci. 44, 1404-1410.

Veverka, J., Wasserman, L. H., Elliot, J., Sagan, C., Liller, W. 1974. The occultation of beta Scorpii by Jupiter. I. The structure of the Jovian upper atmosphere. Astron. J. 79, 73-84.